\documentclass[12pt]{article}
\usepackage{amsmath}
\usepackage{amscd}
\usepackage{amssymb}
\usepackage{array}

\begin{document}
\rightline{CERN-TH/2002-084}

\newcommand{\R}{\mathbb{R}}
\newcommand{\C}{\mathbb{C}}
\newcommand{\Z}{\mathbb{Z}}
\newcommand{\Hb}{\mathbb{H}}

\newcommand{\rE}{\mathrm{E}}
\newcommand{\rSp}{\mathrm{Sp}}
\newcommand{\rSO}{\mathrm{SO}}
\newcommand{\rSL}{\mathrm{SL}}
\newcommand{\rSU}{\mathrm{SU}}
\newcommand{\rUSp}{\mathrm{USp}}
\newcommand{\rU}{\mathrm{U}}
\newcommand{\rF}{\mathrm{F}}
\newcommand{\rGL}{\mathrm{GL}}
\newcommand{\rG}{\mathrm{G}}
\newcommand{\rK}{\mathrm{K}}

\newcommand{\fgl}{\mathfrak{gl}}
\newcommand{\fu}{\mathfrak{u}}
\newcommand{\fsl}{\mathfrak{sl}}
\newcommand{\fsp}{\mathfrak{sp}}
\newcommand{\fusp}{\mathfrak{usp}}
\newcommand{\fsu}{\mathfrak{su}}
\newcommand{\fp}{\mathfrak{p}}
\newcommand{\fso}{\mathfrak{so}}
\newcommand{\fl}{\mathfrak{l}}
\newcommand{\fg}{\mathfrak{g}}
\newcommand{\fr}{\mathfrak{r}}
\newcommand{\fe}{\mathfrak{e}}

\newcommand{\id}{\relax{\rm 1\kern-.35em 1}}
\vskip 1.5cm

  \centerline{\LARGE \bf Duality and  Spontaneously Broken}
\bigskip
 \centerline{\LARGE \bf  Supergravity in Flat Backgrounds}

 \vskip 3cm
\centerline{L. Andrianopoli$^{\sharp\;\natural}$, R.
D'Auria$^{\sharp\;\natural}$, S. Ferrara$^{\sharp\;\flat}$ and M.
A. Lled\'o$^{\sharp\;\natural}$.}

\vskip 1.5cm

\centerline{\it $^\sharp$ Dipartimento di Fisica, Politecnico di
Torino,} \centerline{\it Corso Duca degli Abruzzi 24, I-10129
Torino, Italy. } \medskip

\centerline{\it $\natural$  INFN, Sezione di Torino, Italy. }

\medskip

\centerline{\it $^\flat$ CERN, Theory Division, CH 1211 Geneva 23,
Switzerland, and } \centerline{\it INFN, Laboratori Nazionali di
Frascati, Italy.}

\vskip 1cm

\begin{abstract}
It is shown that the super Higgs mechanism that occurs in a wide
class of models with vanishing cosmological constant (at the
classical level) is obtained by the gauging of a flat group which
must be an electric subgroup of the duality group. If the residual
massive gravitinos which occur in the partial supersymmetry
breaking  are BPS saturated, then the flat group is non abelian.
This is so for all the models obtained by a Scherk-Schwarz
supersymmetry breaking mechanism. If gravitinos occur in long
multiplets, then the flat groups may be abelian. This is the case
of supersymmetry breaking by string compactifications on an
orientifold $\mathrm{T}^6/\Z_2$ with non trivial brane fluxes.
\end{abstract}

\vfill\eject

\section{Introduction}

Spontaneously broken supergravity is a  plausible scenario for a
unified theory of particle interactions below the Planck scale.
Among all the possible models, no-scale supergravities
\cite{cfkn,elnt} have received much attention in the literature
because of the possibility of creating a hierarchy of scales much
lower than the Planck or string scale.

Very recently, no-scale supergravities have emerged from certain
superstring compactifications in presence of brane fluxes
\cite{ps} - \cite{kst}. In these models, the scalar potential and
a non trivial gauging appear as a consequence of the non trivial
flux   along 3-cycles of the internal space. If $\hat
H_{\hat\mu\hat\rho\hat\sigma}$ is a 3-form field strength (the
``$\,\hat{\hbox{ }}\,$" will always denote variables and indices
in the higher dimensional space), then the integral over the
3-cycle is
$$\Phi(x^\mu)=\int_{\mathrm{3-cycle}}dx^{\hat\mu}dx^{\hat\rho}dx^{\hat\sigma}\hat
H_{\hat\mu\hat\rho \hat\sigma}\neq 0.$$

In the Scherk-Schwarz \cite{ss} mechanism, the dimensional
reduction ($D=5\rightarrow D=4$) ansatz  on the 1-form potentials
is \begin{equation}\hat A_{\hat \mu }(x^\nu,x^5)=
e^{Mx^5}A_{\hat\mu}(x_\nu), \qquad \hat\mu, \hat\nu=1,\dots 5,\;\;
\mu,\nu=1,\dots 4.\label{ansatz}\end{equation} $x^5$ is
compactified to a circle $S^1$, and $M$ is a generic element of
the Cartan subalgebra of $\fusp(8)$. The 2-form field strengths
$F^\Lambda_{\hat\mu\hat\nu}$ have also a non trivial flux through
the 1-cycle $S^1$,
\begin{equation}\Phi^\Lambda_\nu=\int_{S^1}(e^{-Mx_5}\hat F)^\Lambda_{\hat\mu\nu}dx^{\hat\mu}=M_\Sigma^\Lambda
A_\nu^\Sigma-\partial_\nu a^\Lambda, \qquad
a^\Lambda=A_5^\Lambda.\label{flux}\end{equation} The combination
in the right hand side of (\ref{flux}) appears in the four
dimensional Lagrangian when making the generalized dimensional
reduction (\ref{ansatz}). Since it is a flux, it must be covariant
under gauge transformations. One can explicitly check it. Under
the infinitesimal transformations \cite{adfl2}
\begin{eqnarray*}&&\delta A_\mu^\Lambda=\partial_\mu\Xi^\Lambda
+M_\Sigma^\Lambda A_\mu^\Sigma\Xi^0+a^\Lambda\partial_\mu\Xi^0\\
&&\delta a^\Lambda=M_\Sigma^\Lambda \Xi^\Sigma +M_\Sigma^\Lambda
a^\Sigma\Xi^0,\end{eqnarray*} the flux transforms as $\delta
\Phi^\Lambda_\mu=\Xi^0M_\Sigma^\Lambda \Phi_\mu^\Sigma$.
\par
For $M=0$ the  term proportional to $A_\mu^\Lambda$ in
(\ref{flux}) is zero. This implies that in the four dimensional
Lagrangian there are no gauge couplings for $A_\mu^\Lambda$, and
the  gauge group of the theory is trivial in the sense that none
of the fields is charged. So an $M\neq 0$ in the dimensional
reduction ansatz is the origin of the gauged gravity in four
dimensions.

We note that the genuine gauge fields in $D=4$ are
$Z_\mu^\Lambda=A_\mu^\Lambda - B_\mu a^\Lambda$ and $B_\mu$, with
$B_\mu$ the vector field coming from the metric (graviphoton). As
shown in Ref. \cite{adfl2}, they are gauge connections of a
non-abelian Lie algebra,  whose generators $\{X_\Lambda , X_0 \}$
obey the commutation relations $$[X_\Lambda , X_0 ] = M_\Lambda
^\Sigma X_\Sigma \, \qquad [X_\Lambda , X_\Sigma ] = 0.$$
$M_\Lambda^\Sigma$ are the matrix elements of $M$ in
(\ref{ansatz}). This is a subalgebra of the U-duality algebra
\footnote{We call U-duality the continuous symmetry of the low
energy effective action that gives rise to the discrete U-duality
symmetry in the full quantum theory.}, which for $N=8$ is
$\fe_{7(7)}$. The subalgebra has dimension 28, and it has a linear
realization on the vector potentials\footnote{We will call such
subalgebras {\it electric} subalgebras.}. This is not a particular
case of any of the gaugings studied in the literature on four
dimensional supergravity \cite{hw,cfgtt}.

A crucial point for the definition of an electric subgroup of the
U-duality group G is the fact that G can be embedded in the
symplectic group $\rSp(2n,\R)$, where $n$ is the number of
electric potentials in the theory.  As we will see, G admits
different symplectic embeddings, which generically are related by
a conjugation with a symplectic transformation. The electric
subgroup is not invariant under this conjugation, then giving rise
to the possibility of non isomorphic maximal electric subgroups.
The standard formulation of supergravity \cite{dwn} (where the
R-symmetry of the theory is manifest) implies the choice of one
particular symplectic embedding. The non standard gaugings
considered here correspond to other choices of the symplectic
embedding which give partial supersymmetry breaking with vanishing
cosmological constant. The symmetry that was electric in the
standard formulation is not fully electric in the other choices.
For example,   in the $N=8$ Scherk--Schwarz supergravity, SO(8)
mixes electric with magnetic field strengths, and so it does
SO(6,$n$) in the case of $N=4$ Scherk--Schwarz supergravity.
Another consequence of the change of a symplectic basis for the
duality group is that the R-symmetry is no longer manifest. In the
above examples, only $\rUSp(8)\subset \rSU(8)$ and
$\rUSp(4)\subset \rSU(4)$ are manifest symmetries.

The paper is organized as follows. In Section \ref{embeddings} we
give some mathematical preliminaries  on electric subgroups of
the U-duality group G. In Section  \ref{N=8} we discuss two
different choices of maximal electric subgroup in $N=8$
supergravity, one giving rise to the standard gaugings and the
other realizing the Scherk--Schwarz mechanism from a purely four
dimensional point of view. In Section \ref{N=4} we consider 3
different choices of electric subalgebras in $N=4$ supergravity,
corresponding to different physical theories and in Section
\ref{N=2} we do a similar analysis for $N=2$ supergravity.

\section{Symplectic embeddings and electric subalgebras \label{embeddings}}

We consider  $D=4$ supergravity  in absence of fluxes. Let G be
the U-duality group of the theory and  $n$ the number of vectors
in the theory, $A_\mu^\Lambda,\; \Lambda=1\dots n$.  The field
strengths $F^\Lambda_{\mu\nu}$ and their duals
$G_\Lambda^{\mu\nu}=\partial\mathcal{L}/\partial
F^\Lambda_{\mu\nu}$ together carry a linear representation of  G.
When G is considered as embedded in $\rSp(2n, \R)$, the
representation carried by $\{F^\Lambda, G_\Lambda\}$ is promoted
to a representation of the full symplectic group \cite{gz}.

An arbitrary matrix of the  Lie algebra $\fsp(2n, \R)$ can be
written in terms of blocks of size $n\times n$
\begin{equation}\label{matrix}X=\begin{pmatrix}a&b\\c&-a^T\end{pmatrix},\qquad
b=b^T, \; c=c^T, \end{equation}
 and $a$ an arbitrary matrix of
$\fgl(n,\R)$. The Lie algebra $\fsp(2n, \R)$ admits then the
decomposition $$\fsp(2n,
\R)=\tilde\fg^0+\tilde\fg^{+1}+\tilde\fg^{-1},$$ with
  $$\tilde\fg^0=\bigr\{\begin{pmatrix}a&0\\0&-a^T\end{pmatrix}\bigl\}, \quad
\tilde\fg^{+1}=\bigr\{\begin{pmatrix}0&0\\c&0\end{pmatrix}\bigl\}
,\quad
\tilde\fg^{-1}=\bigr\{\begin{pmatrix}0&b\\0&0\end{pmatrix}\bigl\}.$$
So $\tilde\fg^0\approx\fgl(n,\R)$, $\tilde\fg^{+1}$ carries the
representation of $\fgl(n,\R)$
$\mathrm{sym}(\mathbf{n'}\otimes\mathbf{n'})$
 and $\tilde\fg^{-1}$  the representation
$\mathrm{sym}(\mathbf{n}\otimes\mathbf{n})$. $\mathbf{n'}$ denotes
the contragradient representation of $\mathbf{n}$. The subalgebra
$\fso(1,1)_z\subset \fgl(n,\R)$ (the subindex $z$ is to
distinguish it from other subalgebras $\fso(1,1)$ that we will
consider in the following), whose generator is the element
$-\frac{1}{2}\id$, acts with charge $+1$ on $\tilde\fg^{+1}$ and
with charge $-1$ on $\tilde\fg^{-1}$ (so the upper indices
indicate this charge). In fact, $\fso(1,1)_z$ defines a grading of
$\fsp(2n, \R)$ and consequently $\tilde\fg^{\pm 1}$ are abelian
subalgebras.
 The matrices of the subalgebra $\tilde\fg^0+\tilde\fg^{+1}$ have lower
block-triangular form ($b=0$). The vector space carrying the
fundamental representation of the symplectic algebra inherits also
a grading and it decomposes as $V=V^+ \oplus V^-$. Notice that
$V^+$ carries a representation of $\tilde\fg^0+\tilde\fg^{+1}$.

We consider now the U-duality group G with Lie algebra $\fg\subset
\fsp(2n)$. Any subalgebra of $\fg$ which is a subalgebra of the
lower triangular matrices $\tilde\fg^0+\tilde\fg^{+1}$ will
transform the field strengths (in $V^+$)  without involving their
magnetic duals. Then, the vector potentials themselves carry a
linear representation of this subalgebra which will be called an
{\it electric subalgebra} of $\fg$ (and generically denoted by
$\fg_{\mathrm{el}}$). The corresponding group,
$\rG_{\mathrm{el}}$, is an {\it electric subgroup} of G. The gauge
group  $\rG_{mathrm{gauge}}$ is a subgroup of the
$\rG_{\mathrm{el}}$ such that its action on the vector potentials
is the adjoint action. As we will see, there is not a unique
maximal electric subgroup of G, and this gives rise to many
different gaugings of the supergravity theory.

In the next sections, we will discuss the examples of $N=8,4,2$
supergravity from this new point of view.

\section{$N=8$ supergravity\label{N=8}}

The U-duality group of $N=8$ supergravity is $\rG=\rE_{7,7}$
\cite{cj}, which can be embedded in several ways in $\rSp(56,\R)$
(there are 28 vector fields). The electric subalgebras will always
be subalgebras of $\fg_{\mathrm{el}}\subset
\tilde\fg^0+\tilde\fg^{+1}$, being in this case
$\tilde\fg^0\approx \fsl(28, \R ) + \fso(1,1)_z$.

\bigskip
 Consider the following
decomposition of $\fe_{7,7}$ \begin{equation}\fe_{7,7}=\fsl(8,\R)
+ \mathbf{70} \label{e77}.\end{equation} $\fsl(8,\R)$ is a maximal
subalgebra of $\fe_{7,7}$ and $ \mathbf{70}$ is an irreducible
representation of $\fsl(8,\R)$, the four-fold antisymmetric.

The representation ${\bf 56}$ of $\fe_{7,7}$ decomposes under the
subgroup $\fsl(8,\R)$ as $$\mathbf{56}\longrightarrow
\mathbf{28}+\mathbf{28'},$$ where $\mathbf{28}$ and $\mathbf{28'}$
are two-fold antisymmetric representations of $\fsl(8,\R)$.

The embedding $\fe_{7,7}\subset \fsp(56,\R)$ is constructed as
follows. We have that $\fsl(8,\R)\subset \fsl(28,\R)$ by means of
the two-fold antisymmetric representation (the $\mathbf{28}$) of
$\fsl(8,\R)$.
 With
a  two-fold antisymmetric tensor we can construct a four-fold
antisymmetric tensor by taking the symmetrized tensor product. In
this way the generators in the
 {\bf 70} of (\ref{e77}) are realized in the two-fold symmetric
 representation of $\fsl(28,\R)$, $\mathrm{sym}(\mathbf{28}\otimes \mathbf{28})$,
 forming the $b$ matrix of (\ref{matrix}),
 $b_{\{[AB][CD]\}}$. Since in $\fsl(8,\R)$ there is
 an invariant, totally antisymmetric tensor $\epsilon_{A_1\dots
 A_8}$, we have another symmetric matrix
 $$c^{\{[A_1A_2][A_3
 A_4]\}}=\frac 1{4!}\epsilon^{A_1\dots
 A_8}b_{\{[A_5A_6][A_7
 A_8]\}}.$$

 This is the standard
embedding of $\fe_{7,7}$ in $\fsp(56,\R)$.  $\fsl(8,\R)$ is a
maximal electric subalgebra. The gauging of different electric
subalgebras of $\fsl(8,\R)$ and of its contractions gives rise to
all the theories described in \cite{hw,cfgtt}. In this choice,
SO(8) is the maximal compact electric subgroup.

\bigskip

 We will now consider a different embedding, which is
the one relevant for the Scherk--Schwarz mechanism.  Consider the
decomposition $$\fe_{7,7}=\fe_{6,6}+\fso(1,1)_{k} +
\mathbf{27_{-2}}+\mathbf{27'_{+2}},$$ where
$\fe_{6,6}+\fso(1,1)_{k}$ is a maximal reductive
subalgebra\footnote{Reductive means that it is the direct sum of a
semisimple algebra plus abelian factors.}. Notice that it is not a
maximal subalgebra.  In five dimensions the U-duality group is
$\rE_{6,6}$ and it is totally electric. If we want to see the four
dimensional theory as the dimensional reduction of a five
dimensional one, this is the natural decomposition to consider,
and the $\fso(1,1)_{k}$ rescales the modulus of the
compactification radius of the 5th dimension. The normalization of
the generator of $\fso(1,1)_{k}$ has been chosen in such way that
the fundamental representation decomposes as
$$\mathbf{56}\rightarrow \mathbf{27_{+1}}
+\mathbf{1_{+3}}+\mathbf{27'_{-1}}+\mathbf{1_{-3}}. $$ (Note that
the ratio one to three of the $\fso(1,1)_{k}$ charges is what one
obtains for the relative charges of the 27 five-dimensional
vectors versus the graviphoton in the standard Kaluza--Klein
reduction).

 We have that $\fe_{6,6}+\fso(1,1)_{k}
\subset\fgl(28,\R)$, by decomposing the fundamental representation
$$\mathbf{28}\rightarrow \mathbf{27_{+1}}+\mathbf{1_{+3}}$$ but
$\fso(1,1)_{\mathrm{k}}$ does not correspond to the trace
generator $\fso(1,1)_z$ in $\fgl(28,\R)$, since all the  vectors
in the representation $\mathbf{28}$  have  the same charge under
$\fso(1,1)_z$, $\mathbf{27_z}+\mathbf{1_z}$. Indeed there is
another subalgebra $\fso(1,1)_r$ in $\fgl(28,\R)$ which commutes
with $\fe_{6,6}$. This comes from the sequence of embeddings
$$\fe_{6,6}\subset\fsl(27,\R)\subset
\fgl(27,\R)=\fsl(27,\R)+\fso(1,1)_r\subset \fsl(28,\R)$$
corresponding to the fact that only 27 of the 28 vectors are
transformed by $\fe_{6,6}$. The charges of the 28 vectors are
$\mathbf{27_r}+\mathbf{1_{-27r}}$. Then $\fso(1,1)_{k}$ turns out
to be a combination of $\fso(1,1)_z$ and $\fso(1,1)_r$.

In this setting the symplectic embedding of $\fe_{7,7}$ in
$\fsp(56,\R)$ is different from the standard one previously
considered. It was explicitly worked out in \cite{adfl2}. Here
$\fg^0 = \fe_{6,6}+\fso(1,1)_{k}$ is the block diagonal part and
$$\fg_{\mathrm{el}} = \fe_{6,6}+\fso(1,1)_{k} +\mathbf{27'_{+2}}$$
is lower block-triangular. Then, it is an electric subalgebra.
\par
Note that in order to have a physical theory the unbroken gauge
group in the $\fg^0$ part must belong to the maximal compact
subgroup of $\fg^0$.

The Scherk--Schwarz \cite{ss,css} mechanism corresponds to the
gauging of an electric subgroup (a ``flat group") with algebra
$\fg_{\mathrm{el}}=\fu(1)\circledS\mathbf{27'_{+2}}$ (semidirect
sum), where
 $\fu(1)$ is a generic element of the Cartan
subalgebra of the maximal compact subgroup $\fusp(8)$ of
$\fe_{6,6}$ \cite{adfl2}. The gauging of this electric group
breaks spontaneously the supersymmetry. Partial breaking is
allowed, and the unbroken supersymmetry algebra has a central
charge.  Central charges are $\fu(1)$ symmetries which belong to
the CSA of G, so if they belong to $\fg_{\mathrm{el}}$ they must
belong to  the maximal compact subalgebra of $\fg^0\subset
\fg_{\mathrm{el}}$. In our case we have one central charge which
is identified with the $\fu(1)$ factor in the semidirect sum
$\fg_{\mathrm{el}}$.

In fact, the Scherk--Schwarz mechanism allows partial breakings
$N=8\rightarrow N'=6,4,2,0$, and the spin 3/2 multiplets are 1/2
BPS, which means that only one central charge is present. The
number of unbroken translational symmetries in the phases
$N'=6,4,2,0$ is, respectively, 15,7,3,3.

\section{$N=4$ supergravity. \label{N=4}}

 As we are going to see, a richer structure emerges in the $N=4$ theory
 because in this case
we can have both non abelian and abelian flat groups, depending on
the particular model we consider.

Let us consider the $N=4$ theory with $n_v+1$ vector multiplets.
The U-duality group is $\rG=\rSO(6,n_v+1)\times\rSL(2,\R)$,
embedded  (in different ways) in  $\rSp(2(6+n_v+1),\R)$, so any
electric subalgebra must have block diagonal part a subalgebra
$\fg^0 \subset \fsl(6+n_v+1,\R) \times \fso(1,1)_z$.

\bigskip

The standard embedding corresponds to take
$$\fso(6,n_v+1)+\fso(1,1)_q\subset \fgl(6+n_v+1,\R).$$
 Here
$\fso(1,1)_q$ is the Cartan subalgebra of $\fsl(2,\R)$ and it is
identified with $\fso(1,1)_z$. The only off-diagonal elements are
the other two generators of $\fsl(2,\R)$, say $X^\pm$. The
electric subalgebra is then the lower triangular subalgebra
$(\fso(6,n_v+1)+\fso(1,1)_q)\circledS \{X^+\}$. This embedding
appears when doing the  compactification of the heterotic string
on $T^6$. (See for example the review of Ref. \cite{gpr}).

\bigskip

We analyze now another symplectic embedding. We take  $n_v=5$ (the
embedding is possible only for $n_v\geq 5$). Then we have the
following decomposition
 $$\fso(6,6)=
\fsl(6,\R)+ \fso(1,1)_s+\mathbf{{15'}^+}+\mathbf{{15}^-}, $$ where
$\mathbf{15}$ is the two-fold antisymmetric representation. Since
$\fsl(n)+\fsl(m)\subset \fsl(nm)$, we have that
$$\fsl(6,\R)+\fsl(2,\R)+\fso(1,1)_s\subset \fgl(12,\R).$$ The
representation $(\mathbf{15'},\mathbf{1})$ is symmetric (the
singlet of $\fsl(2,\R)$ is the two-fold antisymmetric), so we have
that $\mathbf{{15'}^+}\subset\tilde\fg^+\subset \fsp(24,\R)$. This
defines the symplectic embedding. The $\fso(1,1)_s$ is identified
with $\fso(1,1)_z$ of the symplectic algebra.

 The representation $\mathbf{12}$ of
$\fso(6,6)$ decomposes, with respect to $\fsl(6,\R)+\fso(1,1)_z$,
as $$\mathbf{12}\rightarrow \mathbf{6_{+1}}+\mathbf{6_{-1}},$$
thus containing six electric and six magnetic fields, and the
bifundamental of $\fso(6,6)+\fsl(2,\R)$ decomposes as
$$\mathbf{(12,2)}=\mathbf{(6_{+1},2)}_{\mathrm{electric}}+\mathbf{(6_{-1},2)}_{\mathrm{magnetic}}.$$
In particular, we see that $\fsl(2,\R)$ is totally electric.

The twelve vectors gauge an abelian 12-dimensional subgroup of the
$\mathbf{{15'}^+}$ translations.

This model was investigated in  Ref. \cite{tz}, but from our point
of view  it comes from the general analysis of gauging flat
groups. In this case the flat group is completely abelian, since
no central charge is gauged. The theory has four independent mass
parameters, the four masses of the gravitinos. This allows  a
partial supersymmetry breaking without cosmological constant from
$N=4\rightarrow N'=3,2,1,0$, where the massive gravitinos belong
to long (non BPS) massive representations in all the cases, as it
is implied by the fact that the central charge is not gauged and
the fields are not charged under it.
\par
From the analysis of the
consistent truncation of $N=4\rightarrow N'=3$ supergravity, it is
known that $N=4$ supergravity coupled to 6 matter vector
multiplets can indeed be consistently reduced to an $N=3$ theory
coupled to 3 matter multiplets \cite{adfl}. Correspondingly we
have, for the scalar manifolds of such models,
$$\rSU(3,3)/(\rSU(3)\times\rSU(3)\times\rU(1))\subset\rSO(6,6)/(\rSO(6)\times\rSO(6)).$$
In fact,   the Higgs effect in this theory needs the gauging of a
group of dimension 12, spontaneously broken to a group of
dimension 6.
 The
scalar manifolds of the broken phases are
\begin{eqnarray*}& \rSU(3,3)/(\rSU(3)\times\rSU(3)\times\rU(1)),\quad
&\mathrm{for}\quad N'=3\\&
(\rSU(1,1)/\rU(1))\times\rSU(2,2)/(\rSU(2)\times\rSU(2)\times\rU(1)),\quad
&\mathrm{for}\quad N'=2\\& (\rSU(1,1)/\rU(1))^3,\quad
&\mathrm{for}\quad N'=1.\end{eqnarray*} 
A detailed analysis of this model has been given in \cite{dfv}.

 These models are also
analyzed in Ref. \cite{fp,kst} from another point of view. There,
Type IIB supergravity is compactified on the $T_6/\Z_2$
orientifold with brane fluxes turned on and the same pattern of
spontaneous symmetry breaking is found.

\bigskip

 There is still a third symplectic embedding. An $N=4$ theory coupled to
 $n_v +1$ vector multiplets can be obtained by dimensional reduction
of $N=4$ supergravity in $D=5$ with $n_v$ vector multiplets. This
theory is described by a $\sigma$-model $\rSO(1,1) \times
\rSO(5,n_v) /\left(\rSO(5)\times \rSO(n_v)\right)$ \cite{at}.

 In $D=5$, the U-duality
group (totally electric) is $\rSO(5,n_v)\otimes \rSO(1,1)_{d_5}$.
By dimensional reduction from $D=5$ to $D=4$ we know that a
symplectic embedding must exist where the four dimensional
electric group contains $\rSO(5,n_v)\otimes \rSO(1,1)_{d_5}\otimes
\rSO(1,1)_{k}$, where, as in the $N=8$ case, $\rSO(1,1)_{k}$
rescales  the radius of the fifth dimension. We will write
explicitly this embedding later.

The $5+n_v$ vectors $A_\mu^\Lambda$ belonging to the fundamental
representation of $\rSO(5,n_v)$ in $D=5$ and the singlet $C_\mu$
have charges $-1$ and $2$  respectively under $\rSO(1,1)_{d_5}$.
When the dimensional reduction from $D=5$ to $D=4$ is performed we
obtain the extra vector $B_\mu$ coming from the fifth vielbein. It
is neutral under $\rSO(1,1)_{d_5}$ but has charge $3$ with respect
to $\rSO(1,1)_{k}$, while all the other $6+n_v$ five-dimensional
vectors have charge $1$.  Summarizing, the electric and magnetic
field strengths have the following charges $(k,d_5)$ (in a given
normalization) \begin{eqnarray}&&
(\mathbf{5+n_v})_{\mathrm{electric}}^{+1,-1}+\mathbf{1}_{\mathrm{electric}}^{+1,+2}
+\mathbf{1}_{\mathrm{electric}}^{+3,0}\;,\nonumber\\&&
(\mathbf{5+n_v})_{\mathrm{magnetic}}^{-1,+1}+\mathbf{1}_{\mathrm{magnetic}}^{-1,-2}+
\mathbf{1}_{\mathrm{magnetic}}^{-3,0}.\label{pasticcio1}\end{eqnarray}
\par
From a purely four dimensional perspective, the full duality
group is $\rSO(6,n_v+1)\otimes\rSL(2,\R)$. We have the following
decomposition
\begin{eqnarray}&&\fso(6,n_v+1)=\fso(5,n_v)+\fso(1,1)_s+\mathbf{(5+n_v)^+}+\mathbf{(5+n_v)^-}\nonumber\\&&
\fsl(2,\R)=\fso(1,1)_q+\mathrm{span}\{X^+,X^-\}\label{uduality}\end{eqnarray}
where the $(5+n_v)^+$ Peccei-Quinn symmetries appear explicitly.
The representation of the full duality group is the bifundamental,
a doublet of $\fsl(2,\R)$ tensor product with the fundamental of
$\fso(6,n_v+1)$.  We denote by $\fso(1,1)_q\subset\fsl(2,\R)$ the
CSA of $\fsl(2,\R)$. Then the field strengths (electric and
magnetic) have the following charges $(s,q)$
\begin{eqnarray}\bigl(\mathbf{2}, (\mathbf{6+n_v+1})\bigr)&\rightarrow &(\mathbf{5+n_v})^{0,+1/2}+
\mathbf{1}^{-1,+1/2}+\mathbf{1}^{+1,+1/2}+\nonumber\\
&&(\mathbf{5+n_v})^{0,-1/2}+
\mathbf{1}^{-1,-1/2}+\mathbf{1}^{+1,-1/2}.\label{pasticcio2}\end{eqnarray}
In the standard formulation, $\fso(6,n_v+1)$ is fully electric,
but in the symplectic embedding that we are considering this is
not the case.

We consider now the subspace
$$(\mathbf{5+n_v})^{0,+1/2}+\mathbf{1}^{+1,-1/2}+
\mathbf{1}^{+1,+1/2}$$ in (\ref{pasticcio2}). We want to  identify
it with the electric subspace in (\ref{pasticcio1}). Indeed, let
us denote $(X_k,X_{d_5})$ and $(X_s,X_q)$ the generators of the
respective algebras $\fso(1,1)$. From the charges written above we
have that this identification is possible if we assume the
relations
\begin{eqnarray*}\frac 1 2 X_k=X_s+X_q, & X_{d_5}=X_s-2X_q\\
X_s=\frac 1 3 (X_k+X_{d_5}), &X_q=\frac 1 3 (\frac 1 2
X_k-X_{d_5}).\end{eqnarray*} So we see that under this
identification the generators $(X_k,X_{d_5})$ are linear
combinations of $(X_s,X_q)$.

We now give the precise embedding of the U-duality algebra
$\fso(6,n_v+1)\times\fsl(2,\R)$ in $\fsp(2(7+n_v),\R)$.  We
consider first the diagonal subalgebra, $\fgl(7+n_v),\R)$. We have
\begin{eqnarray}
\fgl(7+n_v)&&\simeq \fso(1,1)_z\otimes\fsl(7+n_v) \nonumber
\\  &&\supset
\fso(1,1)_z\otimes \fso(1,1)_r \otimes \fso(1,1)_t\otimes
\fsl(5,n_v)\nonumber\\&&\supset\fso(1,1)_z\otimes \fso(1,1)_r
\otimes \fso(1,1)_t\otimes\fso(5,n_v).\label{gl7}
\end{eqnarray}

Since  $A^\Lambda _\mu, C_\mu, B_\mu$ have to be electric
potentials, we assume they have all the same charge under
$\fso(1,1)_z$. From (\ref{gl7}) we can compute  their charges
$(r,t,z)$ under $\fso(1,1)_r+ \fso(1,1)_t +\fso(1,1)_z$ :
\begin{equation}
(\mathbf{5+n_v})_{(\frac{r} {6+n_v},\frac{t} {5+n_v}, z)}, \quad
\mathbf{1}_{(\frac{r} {6+n_v},-t, z)}; \quad
\mathbf{1}_{(-r,0,z)}.
\end{equation}

It is then straightforward to show that there exist linear
combinations of these three charges which reproduce the charges of
$\fso(1,1)_{d_5}\otimes \fso(1,1)_{k}\subset \fso(6,6)+
\fsl(2,\R)$. We have proved that
$$\fso(5,n_v)+\fso(1,1)_s+\fso(1,1)_q\subset\fgl(7+n_v).$$ The
generators in $(\mathbf{5+n_v})^+$ are embedded in the symplectic
group as (see
 (\ref{matrix}))
\begin{equation*}\label{matriciao}
 a(t)=\begin{pmatrix} 0^{\Lambda}_{\Sigma} & 0^{\Lambda} & t^{\Lambda}\\
 0_{\Sigma} & 0 & 0\\
 0_{\Sigma} & 0 & 0\end{pmatrix},\quad \quad c(t)=\begin{pmatrix}
 0_{\Lambda\Sigma} & t_\Lambda & 0_{\Lambda}\\
 t_\Sigma & 0 & 0\\
 0_{\Sigma} & 0 & 0 \end{pmatrix}, \qquad \Lambda,\Sigma=1,\dots
 5+n_v.
 \end{equation*}
  Note that the difference with the $N=8$ case is that the matrix $c$ here is
 symmetric off-diagonal while for the $\fe_{7(7)}$ case it was symmetric
 diagonal.
The generators $X^\pm$ have also an appropriate embedding in
$\fsp(2(7+n_v),\R)$,  which depends on the linear combination of
$X_r,X_t,X_z$ (generators of the corresponding $\rSO(1,1)$ groups)
which gives $X_q$.
\par
From the symplectic embedding it follows that the transformations
of the electric vectors under the Peccei-Quinn translational
symmetries are:
\begin{equation}\label{transvec}
\delta A^\Lambda_\mu = t^\Lambda B_\mu;\quad \quad \delta
C_\mu=\delta B_\mu =0
\end{equation}
while their magnetic counterparts  have field strengths
transforming as follows:
\begin{eqnarray}\label{transdual}
\delta F^{(m)}_{\Lambda \mu\nu}&=&t_{\Lambda} C_{\mu\nu}
\nonumber \\
\delta C^{(m)}_{\mu\nu}&=&t_{\Lambda} F^{\Lambda}_{\mu\nu} \nonumber \\
\delta B^{(m)}_{\mu\nu}&=&-t^{\Lambda} F^{(m)}_{\Lambda\mu\nu}
\end{eqnarray}
where $t_\Lambda = \eta_{\Lambda\Sigma} t^\Sigma$.

 Note that the charges of $t^\Lambda$ with respect to
 $\fso(1,1)_k$ and $\fso(1,1)_{d_5}$ are $-2$ and $-1$ respectively so
 that the equations (\ref{transvec}) and (\ref{transdual}) respect
 the charge assignements.

 In this symplectic embedding we can gauge flat
groups with algebra $\fg_{\mathrm{el}}=\fu(1)\circledS\fg^+$,
where $\fg^+$ is an abelian subalgebra contained in
$\mathbf{(5+n_v)_+}$. The commutation rules are of the form
$$[X_0,X_\Lambda]=f_{0\Lambda}^{\phantom{0\Lambda}\Sigma}
X_\Sigma, \qquad \Lambda,\Sigma=1,\dots 5+n_v.$$ $X_0$ is a
generic element of the Cartan subalgebra of $\fusp(4)+\fso(n_v)$,
the maximal compact subalgebra of $\fso(5,n_v)$. If $n_v$ is even,
this element depends on $2+n_v/2$ parameters.

Let us assume that $X_0$ is in the CSA of $\fusp(4)$, so it
depends only on two parameters,
$$X_0=\begin{pmatrix}m_1\epsilon&0\\0&m_2\epsilon\end{pmatrix},
\qquad \epsilon=
\begin{pmatrix}0&1\\-1&0\end{pmatrix}.$$
The  model obtained by gauging this algebra  reproduces the
Scherk--Schwarz mechanism applied to $N=4$ supergravity. The
analysis of the explicit symmetry breaking pattern follows the
lines of the $N=8$ supergravity treated in Ref. \cite{css}. The
results are as follows: $m_1$ and $m_2$ are the masses acquired by
the gravitinos and the  gauginos. The non zero eigenvalues of the
Cartan element in the two-fold antisymmetric representation of
$\fusp(4)$ are $\pm(m_1\pm m_2)$, and this gives the masses of the
vectors in the gravity multiplet. At most four vectors can become
massive and four  translational symmetries can be broken.

If $m_1,m_2\neq 0$, then all supersymmetries are broken. If
$m_2=0$, we have a spontaneous breaking $N=4\rightarrow N'=2$. The
massive gravitinos are short, 1/2 BPS multiplets with central
charge given by $X_0$. All the vector multiplets remain massless
and all hypermultiplets acquire a mass $m_1$. The moduli space of
this theory is
$$(\rSU(1,1)/\rU(1))\times\bigr(\rSO(2,n_v+1)/(\rSO(2)\times\rSO(n_v+1))\bigl).$$

\section{$N=2$ supergravity. \label{N=2}}

We would like to discuss now the Scherk--Schwarz mechanism in
$N=2$ supergravity coupled to matter.

Let us first discuss $N=2$ supergravity in $D=5$, with $n_v$
vector multiplets and no  hypermultiplets. The scalars of the
vector multiplets are coordinates of a real manifold of dimension
$n_v$.

The global symmetry of this theory is
 $\rUSp(2) \times \rK$, where $\rUSp(2)$ is the R-symmetry of the $D=5$ theory
 and $\rK$ is the isometry group of the scalar manifold.

In absence of such isometries ($\rK=\id$) then $\rUSp(2)$ is the
only global symmetry which can be used to perform  the
Scherk--Schwarz mechanism. The Cartan generator $X_0$ of
$\rUSp(2)$ is gauged by the graviphoton.

In the theory dimensionally reduced to $D=4$ the manifold
parametrized by the scalars  of the vector multiplets is  a
complex special manifold ${\mathcal M}_v$ of complex dimension
$n_v+1$. We denote by $t^A, \; A=1, \dots n_v+1$ its complex
coordinates. The K\"ahler potential of $\mathcal{M}_v$ is defined
as \cite{dwlvp}
$$\mathcal{K} =-\log \left(d_{ABC} \Im (t^A) \Im (t^B) \Im
(t^C)\right). $$

The scalar manifold  has
 $n_v+1$ translational isometries, however, since the scalars are
 inert under $\rUSp(2)$, these isometries are not gauged.

 Let $m \epsilon$ be a generic element of the CSA of $\rUSp(2)$ with
$$\epsilon = {\begin{pmatrix} 0&1\cr
 -1&0 \end{pmatrix}}.$$
Under the Scherk--Schwarz generalized dimensional
 reduction
  all gravitinos and gauginos, which are $\rUSp(2)$ doublets,
 acquire a mass $e^{{\mathcal K}/2} m$.
All vectors and scalars are massless, so that no gauge symmetry is
broken and the scalar potential vanishes identically.

The Scherk--Schwarz model, in absence of hypermultiplets and with
a special manifold without isometries, corresponds then to $N=2$
supergravity with a Fayet--Iliopoulos term and gives the model
discussed in \cite{ckpdfwg}.

\bigskip

Let us now consider the case in which the special manifold
$\mathcal{M}_v$ has a non trivial isometry group $\rK$. Then one
could consider a  model where the $X_0$ generator has a non zero
component in the Cartan subalgebra of the maximally compact
subgroup of   $\rK$. In this case the gauge algebra would become
non-abelian and some of the vector fields would acquire a mass.

As an example consider the model based on the special manifold
$$\frac{\rSU(3,3)}{\rSU(3)\times \rSU(3)\times \rU(1)}.$$ In the
Scherk--Schwarz mechanism the five dimensional ancestor of this
model is based on the coset $\rSL(3,\C )/\rSU(3)$. Then, in
analogy with the $N=8$ case, we have the decomposition
$$\fsu(3,3)= \fsl(3,\C )+ \fso(1,1)_{k}+ \mathbf{{9'}}^{+2}+
\mathbf{9}^{-2}.$$ The electric subalgebra in $\fsu(3,3)$ is
\cite{adfl} $ \fg_{\mathrm{el}} = \fg^0 + \fg^+$ where
\begin{eqnarray*}
g^0 &=& \fsl(3,\C ) + \fso(1,1)_{k} \nonumber\\ g^+ &=&
\mathbf{9'} = (\mathbf{3',\bar {3'}} ) \mbox{ of } \fsl(3,\C ) .
\end{eqnarray*}
The vector field strengths and their magnetic duals are in the
$\mathbf{20}$ (three-fold antisymmetric) of $\fsu(3,3)$
\cite{ffs}. This representation decomposes as follows under
$\fsl(3,\C) + \fso(1,1)_k$ $$\mathbf{20} \rightarrow
\mathbf{9}^{+1} +\mathbf{1}^{+3} + \mathbf{9^\prime}^{ -1}
+\mathbf{1}^{-3}$$

The gauge algebra is therefore $X_0\circledS \fg^+$ where $X_0$ is
a Cartan element of $\rSU(3) \subset \rSL(3,\C )$, under which
$g^+$ is in the $(\mathbf{8+1})$ representation.

The 10 dimensional Lie algebra $$ [X_\Lambda , X_0 ] =
M_\Lambda^\Sigma X_\Sigma \, ; \qquad [X_\Lambda , X_\Sigma ] =0
$$ has structure constants $M_\Lambda^\Sigma$ given by a diagonal
matrix with entries
\begin{equation}
\left\{ \pm (a_1 - a_2 ), \pm (2a_1 + a_2 ), \pm (a_1 +2 a_2 ),
0,0,0\right\}.\label{eigenvalues}
\end{equation}
In this case the central charge does not break the supersymmetry,
but the nine vector multiplets will organize in three massive
1/2-BPS multiplets with masses proportional to $|a_1-a_2 |$,
$|2a_1+a_2 |$, $|a_1+2a_2 |$, and three massless ones, the fourth
massless vector being the graviphoton.

In the broken phase, the moduli space of this theory is
\cite{adf,adfl}
\begin{equation}\left(\frac{SU(1,1)}{U(1)}\right)^3.\label{km}\end{equation} The mass matrix is
\begin{equation*}
\tilde M_\Lambda^\Sigma =e^{\mathcal{K}/2}M_\Lambda^\Sigma.
\end{equation*}
where  $\mathcal{K}$ is the K\"ahler potential of the manifold
(\ref{km}). There are six broken symmetries, corresponding to the
difference ($9-3=6$) in the number of translational symmetries
between the unbroken and broken phase. Since the masses of the
gravitinos vanish, the scalar potential is positive definite and
it vanishes at the minimum.

This model can be easily generalized by taking $X_0$ to be a
linear combination of a Cartan element of $\rSU(3)$ with the one
of $\rUSp(2)$ (with eigenvalue $m$). In this case the gravitinos
get a mass $m$ (so that the supersymmetry is completely broken),
vectors and scalars still have masses given by the eigenvalues
(\ref{eigenvalues}) of $M_\Lambda^\Sigma$, while the gauginos have
masses shifted by $m$ with respect to the bosons.

\bigskip

Let us briefly discuss the Scherk--Schwarz mechanism when also
hypermultiplets are present. In this case, since the hyperscalars
transform under $\rUSp(2)$, to perform the gauging of $X_0$ in the
CSA of $\rUSp(2)$ it is  required that the quaternionic manifold
${\mathcal M}_Q$ parametrized  by them has at least an $\rUSp(2)$
isometry. This would be for instance the case for all the
quaternionic symmetric spaces \cite{gst,bw,cvp} but would not be
true for more general manifolds as for instance the general
quaternionic spaces obtained through c-map \cite{cfg,fs}.

If the quaternionic manifold ${\mathcal M}_Q$ has some larger
isometry $\rK'$, we may use $\rK'$ for the Scherk--Schwarz
mechanism.

In the case where only $\rUSp(2)$ is used as a global symmetry,
then the hyperscalars would acquire a mass set by $m$, while the
hyperinos would remain massless. Then, a scalar potential would
appear.

\bigskip

The Scherk--Schwarz mechanism allows models either with the $N=2$
supersymmetry left unbroken, or completely broken to $N=0$.

To obtain partial breaking $N=2 \to N=1$ we need hypermultiplets
to be present and to gauge at least two translational isometries
of the quaternionic manifold. The minimal model \cite{cgp,fgp} has
one vector multiplet whose complex scalar parametrizes the special
manifold
\begin{equation}{\cal M}_v =\frac{\rSU(1,1)}{\rU(1)}
\end{equation}
with K\"ahler potential $\mathcal{K} = - \log (z+\bar z)$, and one
hypermultiplet whose scalars parametrize  the quaternionic
manifold
\begin{equation}
{\cal M}_Q = \frac{\rUSp(2,2)}{\rUSp(2)\times \rUSp(2)}.
\end{equation}
This contains three axions with  charges  $(g_1,g_2)$ with respect
to the two gauge fields $(B_\mu , Z_\mu )$.

Since ${\cal M}_Q$ has three translational isometries, the
unbroken $N=1$ theory has still one translational isometry coming
from the quaternionic manifold. The moduli space of the $N=1$
theory is $${\cal M}_v \times
\left(\frac{\rSU(1,1)}{\rU(1)}\right)_Q.$$ $(\rSU(1,1)/\rU(1))_Q$
is a sub-manifold of $\mathcal{M}_Q$ and the spectrum contains a
massive gravitino multiplet and two massless chiral multiplets.
Since no compact generators are gauged, the gauge algebra is
purely abelian and this model is similar to the $N=4$ one with
abelian gauging of the translational isometries considered in
Section \ref{N=4} \cite{tz}. As in that model, the two gravitinos
acquire masses
\begin{equation}
m_{1,2}= |g_1 \pm g_2 | e^{\tilde{\cal K}/2}
\end{equation}
where $\tilde{\cal K}$ is the K\"ahler potential of
$$\left(\frac{SU(1,1)}{U(1)}\right)_v \times
\left(\frac{SU(1,1)}{U(1)}\right)_Q.$$ An $N=1$ theory is found
for $|g_1| =|g_2|$.

This model has vanishing potential since the only scalars which
become massive are the two Goldstone bosons of the $\R^2$ gauged
isometry, which are eaten by the vectors that become massive.

Note that the number of massive vectors is given by $4-2=2$, that
is the difference in number between the translational isometries
of the $N=2$ scalar manifolds and the corresponding ones in the
$N=1$ moduli space. This agrees with the general analysis of
\cite{lo}. A possible generalization of this model to many hyper
and vector multiplets has been given in \cite{fgpt}.

\section{Summary}
In this paper we have explored new gaugings of extended supergravity in which different maximal subgroups
of the duality group have a diagonal embedding in the electric-magnetic symplectic duality rotations.
\par
The gauging of ``flat groups'' allows to obtain a class of models which are the extension of no-scale
supergravity to higher N-theories.
These models encompass the Scherk--Schwarz generalized dimensional reduction and supergravity or M-theory 
compactifications in presence of brane-fluxes.
\par
Among these models, of particular interest for phenomenological studies are the $N=2 \to N=1$ models, which can be
obtained, for example, in IIB orientifolds with three-form fluxes turned on. 
\par
Finally, another possibility of obtaining an $N=2 \to N=1$
breaking is to compactify a five dimensional theory on a $S_1 /
\Z_2$ orientifold \`a la Scherk--Schwarz, by introducing a $\Z_2$
parity which truncates the second gravitino \cite{mp} -
\cite{gq2}. This is the mechanism used  in the supergravity
version of the Randall-Sundrum scenario and will be discussed
elsewhere.

\section*{Acknowledgements}

S. F. would like to thank the Dipartimento di Fisica, Politecnico
di Torino for its kind hospitality during the  completion of this
work.   Work supported in part by the European Community's Human
Potential Program under contract HPRN-CT-2000-00131 Quantum
Space-Time, in which L. A.,  R. D. and M. A. Ll. are associated to
Torino University. The work of S. F. has also  been supported by
the D.O.E. grant DE-FG03-91ER40662, Task C.

\end{document}